\documentclass[reprint,amsmath,amssymb,aps,twocolumn,superscriptaddress]{revtex4-1}
\usepackage[utf8]{inputenc}
\usepackage{amsmath,amssymb}
\usepackage{amsthm,amscd,amsfonts}
\usepackage{dcolumn}
\usepackage[dvipsnames]{xcolor}

\newcommand{\G}{\mathcal{G}}
\usepackage[colorinlistoftodos,textsize=footnotesize]{todonotes}

\begin{document}
\preprint{APS/123-QED}

\title{Generalized Spontaneous Symmetry Breaking}

\author{Sin\'ead M. Griffin}
\email{sgriffin@lbl.gov}
\affiliation{Materials Sciences Division, Lawrence Berkeley National Laboratory, Berkeley, CA 94720, USA}
\affiliation{The Molecular Foundry, Lawrence Berkeley National Laboratory, Berkeley, CA 94720, USA}

\author{Michele Schiavina}
\email{micschia@ethz.ch}
\affiliation{ETH Z\"urich, Department of Mathematics, R\"amistrasse 101, 8092 Z\"urich, Switzerland,}
\affiliation{Institute for Theoretical Physics, Wolfgang-Pauli Strasse 27, 8093, Z\"urich, Switzerland}


\begin{abstract}
    We propose a new type of symmetry breaking mechanism that takes into account boundary defects of a model, and show how it can detect surface modes by interpreting them as the order parameter associated with the generalized symmetry breaking. We argue that this mechanism is analogous to and yet distinct from the Ginzburg--Landau, and show how the two communicate through the specification of boundary conditions.
\end{abstract}

\maketitle

\section{Introduction}
Much of modern condensed matter physics has been concerned with bulk, equilibrium properties of systems, which fit into the Ginzburg--Landau (GL) paradigm of symmetry-breaking phase transitions \cite{Landau/Ginzburg:1950}.   These include conventional magnetic, ferroelectric and superconducting orders, and more complicated exotic and/or multi-order ground states including high-T$_C$ superconductors and multiferroics \cite{Frank/Lemm:2016,Salje/Carpenter:2011, Griffin/Spaldin:2017}.  A key tenet of GL theory is the identification of an ``order parameter'', an observable quantity whose onset under varying an external parameter beyond a critical value detects a change in the global behavior of the system, often described through its symmetry profile. This is known as spontaneous symmetry breaking (SSB), and the ideas underlying this description have been successfully extended to gauge theories \cite{Goldstone:1962,Kibble:2015}.

On the other hand, the discovery of topological phenomena like the fractional Quantum Hall effect and topological phases materials such as topological insulators, semimetals and superconductors has challenged the universal applicability of the GL framework to phase transitions in condensed matter physics where the order parameter is nonlocal \cite{Moore:2010,Bernevig/Hughes:2013}.  In these instances, the underlying topology of the physical system becomes relevant, going beyond the local order parameter descriptor needed for a GL field-theoretic picture.

So far, the description of such topological phases has been accomplished by means of topological field theories, which represent global, effective excitation of microscopic states and encompass the topological properties of the underlying spacetime manifold. Examples include the well-known Chern--Simons theory description of Quantum Hall systems, or the link between topological insulator states and topological $BF$ theory \cite{Frohlich/Zee:1991, Frohlich:1994vq, ChoMoore}. These topological field theories share a common feature: they present a rich symmetry structure that arises when boundaries are included in the description, often resulting in the emergence of surface modes. This peculiar non-local bulk-boundary correspondence can result in dramatically different properties of the system's edges and its bulk. This has observable consequences in materials: for example, topological insulators are insulating in the bulk but have metallic surface states \cite{Hsieh_et_al:2008}. However, consensus on an order-parameter-type description of such non-local properties is still lacking, although it would serve as invaluable conceptual tool for theoretical modeling and prediction.

Order parameters for GL phase transitions are governed by the critical behavior of an effective field theory, where one can equivalently describe these transitions by comparing symmetries of vacua and overall symmetries of the model.
In this Letter, we generalize this and propose a new framework to interpret the emergence of topological order in physical systems as a generalized symmetry breaking paradigm.  Rather than comparing the symmetries of vacua with the group of symmetries of the model, we propose a comparison of the symmetry groups ``before and after'' introducing a boundary (or defect) in the system. The presence of boundaries --- which can be controlled by external parameters --- can reduce the overall symmetry enjoyed by the system. It is therefore natural to look for observables that can detect said symmetry reduction such as edge-modes or surface states, which can be experimentally observed.

From this perspective, we can consider both GL phase transitions and the emergence of topological phases as instances of a unified symmetry breaking paradigm. Here we propose the notion of \textit{generalized spontaneous symmetry breaking}, which is detected by a change of the symmetry profile between two effective models, before and after passing a critical point. A change of the symmetry properties of solutions of the Euler--Lagrange equations  of the associated Lagrangian model detects a GL phase transition, while a change in the overall intrinsic symmetry, induced by a boundary, detects the onset of a topological phase. Crucially, both can happen simultaneously. Surface states become then natural markers of such onset, analogous to the usual order parameters of GL models.

In Section \ref{s:GL} we rephrase the GL-SSB construction so that our generalization to the boundary/defect scenario (Section \ref{s:bound}) becomes natural. Moreover, we show how this extended notion of symmetry breaking interacts with usual SSB scenarios through boundary conditions, and provides an overarching framework to discuss both simultaneously.

\section{Spontaneous symmetry breaking in  Ginzburg-Landau Theory}\label{s:GL}

Spontaneous symmetry breaking phase transitions are detected when the ground state of a system no longer possess the full symmetry of its effective Lagrangian. For a model with symmetry group $G$ that has an SSB phase transition, the broken ground state symmetry will be described by a subgroup $H$ of $G$.  In phase transitions where there are more than one vacuum choice,  passing through the critical point forces the system to choose one, and it becomes `spontaneously broken'. We summarize here the description of SSBs in terms of quotients of symmetry groups, and give examples where this is applied to both global and gauge theories. We begin by a step-by-step break down of  spontaneous symmetry breaking (SSB) \cite{PS,Zinn:2002}: 
\begin{enumerate}
    \item We describe two different phases of a system by two Lagrangian densities $L_1,L_2$. Often $L_2$ is obtained from $L_1$ via a smooth family of Lagrangians $L_t$ that changes characteristics at a given value of the parameter $T=T_C$. 
    The (global) symmetry group of the theory is the same, i.e. both $L_1$ and $L_2$ are invariant under the action of $G$.

    \item We now ask whether the critical points of the Lagrangians $L_{1,2}$ enjoy the same $G$-symmetry. SSB is associated with minima of the potential $V$, which are (constant) critical points $\phi^{(i)}_0$ of $L_i$, enjoying a smaller symmetry. We encode this by saying that their stabilising subgroup is properly contained in $G$, i.e. $H_{\phi_0^{(i)}}\subsetneq G$.

    \item Lastly, we can choose to expand around one such minimum and interpret certain fluctuations as symmetry-broken degrees of freedom -- also referred to as Goldstone bosons. The perturbed action functional is still $G$-invariant, although possibly non-manifestly so.
\end{enumerate}
The whole set of critical points of a Lagrangian $L_i$ will split into orbits --- the set of solutions that can be reached from $\phi_0^{(i)}$ by means of a group action --- 
 which can be described as the cosets $G/H_{\phi_0^{(i)}}$.

\subsection{Global symmetry breaking systems}
Landau theory prescribes a phenomenological description of phase transitions using the concept of an \textit{order parameter}, $\phi$. Detecting a difference in the stablizing subgroups associated to different critical points (across phases) allows us to define an order parameter as a coordinate in the orbit of a critical point with smaller stabilising subgroup. In other words, the set of points on this orbit is usually taken as a definition of the order parameter\cite{Nash/Sen:1988}. For our purposes, it is more convenient to instead consider the order parameter \emph{space}: the set of possible values of the parameter\cite{Sethna:1992}.

We present an example to make these ideas concrete: Consider a field $\phi:M\to \mathbb{R}^n$ on a space-time manifold $M$, with quartic potential $V(\phi) = a\|\phi\|^2 + b\|\phi\|^4$, invariant under $O(n)$-action. We can identify two different phases of the system depending on the parameters $a,b\in \mathbb{R}$. When $V_1$ has the shape of a single well, the Euler--Lagrange equations have only one constant solution  ($\phi_0\equiv0$), which is invariant under the full group action ($H_{\phi_0}=O(n)$). Here, the order parameter (space) collapses to 
a single point because $H_{\phi_0}=G$, and the only orbit is parametrised by the constant vacuum solution $\phi_0=0$. This is referred to as the high symmetry phase. For a quartic potential $V_2$, a given constant solution will only enjoy a residual $O(n-1)$ symmetry. The order parameter space (for constant solutions) is the coset $O(n)/O(n-1)\simeq S^{n-1}$, and every solution $\phi_0$ can be obtained by means of an $O(n)$ rotation from a reference one. We say that the order parameter acquires a nonzero value because it is identified with (a point in) the orbit of a solution $\phi_0$ with a smaller symmetry group. This is a low symmetry phase.

This traditional approach works very well when interested in vacua/ground states of the system, i.e. constant solutions of the Euler--Lagrange equations of the model. However, the orbit interpretation we describe above can be extended to critical points that are more general than vacua, including local maxima of the potential, nonconstant solutions, etc. A unified description of the order parameter \emph{space} is available through the union of cosets:
\begin{equation}\label{e:OP}
    \mathcal{OP}(L_i) := \bigcup_{\phi_0\in \mathrm{EL}(L_i)}G/H_{\phi_0},
\end{equation}
where $\mathrm{EL}(L_i)$ denotes the set of solutions of the Euler--Lagrange (EL) equations for the Lagrangian $L_i$.

Typically, the onset of a SSB phase transition is encoded in the order parameter acquiring a nonzero expectation value between one phase and another, which can be detected in experiment. By virtue of the reformulation of \eqref{e:OP}, we can now detect a phase transition by observing changes in the orbits' structure between one phase and the other. Notice that in definition \eqref{e:OP} we discard neither local maxima of the action functional, nor local (but not global) minima. This opens up a number of subtle scenarios, where the symmetry breaking might be detected by a different behaviour of critical points other than global minima of the system. It has the advantage of including, e.g., changes in the metastable states \cite{VodPeeMeta, Huang_et_al:2014} as well as vortex solutions and topological defects (see \cite{HiranoRybicki,AftDan,Zurek:1996,Kibble:1976,SERFATYVortices,bethuel1994ginzburg}).

\subsection{Local symmetry breaking systems}\label{s:gaugeSSB}
The construction outlined above is readily extended to models that admit a local symmetry group such as gauge theories, where one looks at $\mathcal{G}=C^\infty(M,G)$.

Once again, critical points $\phi_0$ of the action functional might have smaller stabilisers than the whole group: $\mathcal{H}_{\phi_0}\subsetneq\mathcal{G}$. Then, as in the global case, we can look at the orbit $\mathcal{O}_{\phi_0}$ and identify it with the coset $\mathcal{G}/\mathcal{H}_{\phi_0}$. We define the \emph{order parameter space} of the i\textsuperscript{th} phase as:
\begin{equation}\label{e:OPgauge}
    \mathcal{OP}(L_i):= \bigcup_{\phi_0\in \mathrm{EL}(L_i)}\mathcal{G}/\mathcal{H}_{\phi_0}
\end{equation}

The standard example in this class is a gauge field coupled to a Higgs scalar, with a potential term that admits nontrivial (global) minima. This scenario is described by a group reduction of $\G$ to its fiberwise subgroup $\mathcal{H}_{\phi_0}=C^\infty(M,H_{\phi_0})$, where $H_{\phi_0}$ is the stabiliser of the (local) minimum of the associated GL problem. However, besides the sector of constant solutions to the EL equations, the order parameter space of Equation \eqref{e:OPgauge} contains information about more complex phases, including non-topological solitonic solutions\cite{Nugaev/Shkerin:2020}. Moreover, there are many ways in which one can construct a nontrivial stabilizer in (the $\infty$-dimensional group) $\mathcal{G}$, which greatly increases the number of possible, nontrivial, symmetry breaking scenarios \cite{GarlandMurray}.

We note that this picture is independent of any interpretation issues related to the nature of Goldstone modes. The gauge field becomes massive simply as a result of a change in the orbit structure of $\mathcal{OP}$, i.e. the existence of other nontrivial critical points of $S$. This enables us to expand around a nontrivial solution and discuss the properties of the fluctuations.

\section{Symmetry breaking \emph{via} boundaries}\label{s:bound}
In this section we propose an extension of the previous prescription to the case of a field theory on a manifold that admits a nontrivial boundary. We consider here a loose definition of boundaries, to include the case of system interfaces or defects.  Our notion does not distinguish a boundary from a given hypersurface, and effectively we speak of regions in spacetime on which the behaviour of field configurations can be specified. This is flexible enough to be applicable to all general instances considered here.

When a boundary is present, a number of considerations arise that interfere with the discussion of Section \ref{s:GL}. Crucially, we first need to understand which symmetry groups  undergo a spontaneous breaking, as this will differ from the previous scenario. Then, we address how boundary conditions are included in our prescription, and how they interact with symmetry breaking.

\subsection{Intrinsic and residual symmetry of the model}
The space of intrinsic symmetries of a field theory depends uniquely on the structure needed to define the model. For a gauge theory, where a principal bundle is the main datum \footnote{More abstractly this can be taken to be the group of automorphism of the fiber bundle on which the theory is defined.}, this is typically  $\G=C^\infty(M,G)$. 

Within $\G$, we identify the true symmetries of the action functional $\G(S)=\{g\in\G | g\cdot S=S$\}. As we will see, the introduction of a boundary might yield $\G\supsetneq\G(S)$, effectively breaking down the (intrinsic) symmetry. It is useful at this stage to define transformations that leave the fields fixed at the boundary/interface: the subgroup $\G_0=\{g\in\G\ |\ g\vert_{\partial M}=\mathrm{id}\}\subset \G$. We call the resulting quotient $\G^\partial:= \G/\G_0$ the residual symmetry group on the boundary \footnote{If $\G_0$ is not a normal subgroup, the quotient might have the structure of a groupoid. It is a group for the standard case of gauge theory. See \cite[Sect. 11.6]{BW} for the remarkable example of General Relativity.}, which for gauge theories is  $\G^\partial=C^\infty(\partial M, G)$.

We next present a few examples to clarify these points. Consider once again the case of a complex scalar field coupled to a $U(n)$ gauge field (Yang--Mills theory)
\begin{equation}\label{e:YMmatter}
    S= \int \mathrm{tr}((D\phi)_\mu (D\phi)^\mu  - \frac14 F_{\mu\nu}F^{\mu\nu}).
\end{equation}
Crucially, this action functional is invariant with respect to the full group of intrinsic symmetries $\G$ even in the presence of boundaries, i.e. $\G(S)=\G$.

A theory that instead presents a mismatch between $\G$ and $\G(S)$ is Chern Simons theory:
\begin{equation}
    S_{CS}=\int\mathrm{tr}\left[ \frac12 AdA + \frac16 A[A,A]\right].
\end{equation}
In the presence of a boundary, $S_{CS}$ is not invariant under gauge transformations and the group of symmetries that preserve this action \footnote{To be precise one should look at $\exp(\frac{i}{\hbar}kS_{CS})$ with $k$ integer. Indeed the Chern--Simons action is not a function(al) but rather a section of a line bundle.} is $\G(S_{CS})=\G_0$. Analogously, the so-called $BF$ model also presents a mismatch: in dimension 3 (and higher), the action functional
\begin{equation}
    S_{BF}=\int \mathrm{tr}\left[B F_A\right]
\end{equation}
has a symmetry structure such that $\G\supsetneq \G(S_{BF})\supsetneq \G_0$.

Let us briefly comment on this: while the symmetries of Yang-Mills theory (with scalar matter) coincide with the geometric symmetries given by the full gauge group $\G$ even in the presence of a boundary, the symmetries of Chern--Simons and BF theories are effectively reduced to $\G(S_{CS})=\G_0$ and $\G(S_{BF})$ respectively. This phenomenon has an abstract mathematical explanation, but we would like to propose a physical interpretation.

It is known that Chern--Simons and $BF$ theories describe \emph{topologically protected} phases of matter: they provide effective descriptions of Quantum Hall systems and topological insulators \cite{ChoMoore}, which possess bulk-boundary phenomena manifesting in symmetry-protected edge states at their boundaries\cite{Frohlich/Zee:1991, ChoMoore}. The algebra of edge currents is tightly related to the dynamical properties of chiral models, which can be seen as boundary field theories precisely for group-valued fields in $\G/\G(S_{CS})\simeq C^\infty(\partial M,G)$ (analogously for $BF$, see \cite{MSW} for a modern analysis). On the other hand, there are no known topological phases of matter that admit an effective Yang--Mills description, and YM is not believed to give rise to edge currents akin to the ones that emerge in Chern--Simons. We therefore argue that the very presence of a boundary can be considered as a generalized mechanism of symmetry breaking, which instead of comparing the symmetries of given solutions $\phi_0$, compares the effective symmetry of a system with defects against its \emph{intrinsic} symmetry datum, which is fixed by the geometry of the effective field theory used to describe the system.

In light of these observations, we propose the notion of an \emph{intrinsic order parameter space} for an action $S$:
\begin{equation}
    \mathcal{IOP}(S) := \mathcal{G}/\mathcal{G}(S).
\end{equation}
We have observed how it can be used to detect the emergence of edge phenomena, which we wish to interpret as symmetry breaking triggered by the presence of boundaries. This is  similar in spirit to how changes of the order parameter parameter space detects GL spontaneous symmetry breaking. The consequence of this is the interpretation of edge or surface modes as geometric order parameters for topological states of matter, bringing different types of symmetry breaking under the same overarching framework.

Note that, in General Relativity, the group of $4d$-diffeomorphisms that preserve the boundary (think of a Cauchy surface) is not a normal subgroup of the group of all diffeomorphisms, and the quotient $\mathcal{IOP}(S_{GR})=\G/\G_0$ has the structure of a ``groupoid'' \cite{BW} (suggesting that the algebraic structure of currents might be highly nontrivial). 
We believe that interpreting phenomena related to the specification of a boundary/hypersurface in terms of generalised symmetry breaking could provide a starting point in the understanding holographic phenomena and their relation with condensed matter systems.

\subsection{Boundary conditions}
The second immediate consideration that arises when the effective model admits a boundary is the need for boundary conditions. Specifically, we argue that the choice of a boundary condition plays a role in symmetry breaking. 
Looking at the model in \eqref{e:YMmatter}, we can conceive a boundary term 
\begin{equation}\label{Kboundary}
S_K = S + \int_{\partial M}\langle{\phi},K\phi\rangle
\end{equation}
with $K$ a matrix \footnote{Technically a boundary volume form with values in the adjoint representation $\mathfrak{g}$}. Adding this term to the action specifies a boundary condition (see e.g. \cite{PonsBoundary}), and it may change the symmetry group $\G(S)\to\G(S_K)\subseteq \G$.

So far we have not discussed the symmetry of solutions of the Euler--Lagrange equations in this intrinsic setup. However, $K$ might be invariant under conjugation with some group elements, and this gives a bound on the symmetry that a solution with $K$-dependent boundary condition is allowed to enjoy \cite{Goard} \footnote{Although one would be led to think that the symmetry of the boundary condition is the maximum symmetry attainable by a solution with said boundary conditions, examples exists where this can be relaxed \cite{Goard}.}.

Hence, on top of the intrinsic symmetry breaking scenario that results from having boundaries, we can interact with the symmetry breaking mechanism outlined in Section \ref{s:GL}. To do this, start with a $K$-dependent boundary condition, so that the overall gauge group is given by $\G(S_K)$. Let us then pick $\phi_0\in\mathrm{EL}(S_K)$, a field that solves the Euler--Lagrange equations with the given boundary condition. This will have stabiliser $\mathcal{H}_{\phi_0}\subset \G(S_K)$, and the order parameter space will be 
\begin{equation}
    \mathcal{{OP}}(S_K):= \bigcup_{\phi_0\in\mathrm{EL}(S_K)}\G(S_K)/\mathcal{H}_{\phi_0}.
\end{equation}
An intrinsic symmetry breaking is triggered when $\G(S_K)\subsetneq\G$. If, additionally, $\mathcal{H}_{\phi_0}\subsetneq \G(S_K)$ we have a further group reduction and both intrinsic and GL symmetry breaking are manifest.

In Chern--Simons theory, we can fix a (holomorphic) boundary condition by decomposing $A$ as \footnote{This requires the introduction of a metric or a complex structure on $\partial M$.} $A\vert_{\partial M} = A^{1,0} dz + A^{0,1} d\bar{z}$, and adding a boundary term: 
\begin{equation}\label{holomorphicCS}
S_{CS}^{1,0}=S_{CS} + \int_{\partial M} A^{1,0}A^{0,1}. 
\end{equation}
The net result is that of changing the space of symmetries of the theory to yet another group $\G_0\subset \G(S_{CS})\subset\G(S^{1,0}_{CS}) \subset \G$. Notice that while the addition of the boundary term in Yang--Mills likely reduces the overall symmetry of the system, the holomorphic boundary condition \eqref{holomorphicCS} for Chern--Simons increases it.

\section{Outlook and further research}
In this Letter, we extended the notion of symmetry breaking to the case of an (effective) field theory on a manifold with boundary or defects. This introduces the concept of intrinsic symmetries of a model, as an \emph{a-priori} symmetry descriptor that may be broken down when a boundary is considered, as in the cases of topological Chern--Simons and $BF$ theories. This prescription is embedded with the idea of an order parameter for this type of symmetry breaking, which is associated with edge and surface modes.

Moreover, the intrinsic symmetry breaking mechanism we have outlined is a natural extension of the usual SSB scenarios, through the specification of boundary conditions, which can alter both the intrinsic symmetry and symmetries of vacua/critical points. This unified prescription suggests an overarching framework of both GL and topological phase transitions, which up until now have been considered examples of unrelated paradigms.

The main feature of SSB scenarios is a comparison of symmetry profiles of a model before and after a critical point. This general paradigm can be used to describe GL phase transitions, or the onset of topological phases, according to the type of symmetries one considers. While GL SSB is detected by a comparison of symmetries of solutions of the Euler-Lagrange equations, topological phases appear to be detected by the breakdown of intrinsic symmetries of a model, triggered by the presence of a boundary/defect. 

Importantly, both symmetry-breaking mechanisms are tunable. Indeed, we conclude this Letter by pointing out that a boundary can be introduced/controlled in a solid state system by means of external parameters, for example interfaces/defects can be introduced by chemical and physical processes. In systems where the introduction of such boundaries induces a breakdown of the intrinsic symmetries describing the system, our proposed generalized SSB diagnoses the presence of boundary/edge modes.

While completing this work, we became aware of \cite{ElliottGwilliam}. The mathematical techniques therein are compatible with this work, and a bridging link is given by the boundary analysis of field theory presented in \cite{MSW}.

\section{Acknowledgements}
We thank N.\ Reshetikhin, N.\ Beisert, A.\ S.\ Cattaneo, J. Vinson and T. Smidt for helpful discussions. S.M.G. was supported by the Laboratory Directed Research and Development Program of LBNL under the DoE Contract No. DE-AC02-05CH11231. Work performed at the Molecular Foundry was supported by the Office of Science, Office of Basic Energy Sciences, of the U.S. Department of Energy under the same contract No.

This research was (partly) supported by the NCCR SwissMAP, funded by the Swiss National Science Foundation. M.S. acknowledges partial support from Swiss National Science Foundation grants P2ZHP2\_164999 and P300P2\_177862.

\bibliography{refs}
\end{document}